\documentclass[showpacs,showkeys,eprintnumbers,amssymb,amsmath]{revtex4}

\usepackage{amsmath}
\usepackage{amssymb}
\usepackage{eucal}
\usepackage{graphicx}
\newcommand{\be}{\begin{equation}}
\newcommand{\bse}{\begin{subequations}}
\newcommand{\ese}{\end{subequations}}
\newcommand{\ba}{\begin{eqnarray}}
\newcommand{\ea}{\end{eqnarray}}
\newcommand{\ee}{\end{equation}}

\begin{document}

\vspace*{2mm}

\title{ Measuring CP-violating phases through studying the polarization
of the final particles in $\mu \to eee$  }

\author{Yasaman Farzan}
\email{yasaman@theory.ipm.ac.ir}

\affiliation{ School of physics, Institute for Research in
Fundamental Sciences (IPM), P.O. Box 19395-5531, Tehran, Iran\\}

\begin{abstract}
It is shown that   the polarizations of the final particles in
$\mu^+\to e^+e^-e^+$ provide us with  information on the
CP-violating phases  of the effective Lagrangian leading to this
Lepton Flavor Violating (LFV) decay.

\end{abstract}

\date{\today}
\maketitle
\section{Introduction}
In the context of the ``old" Standard Model (SM) with zero
neutrino masses, the lepton flavor is conserved. As a result, the
LFV decays such as $\mu \to eee$ and $\mu \to e \gamma$ are
strictly forbidden within the ``old" SM. Within the ``new" SM with
sources of LFV in the neutrino mass matrix, such decays are in
principle allowed but their rates are suppressed by powers of the
neutrino mass and are therefore beyond the reach of any searches
in the foreseeable future \cite{petcov}. However, a variety of
beyond SM scenarios can lead to rates for these processes
exceeding the present experimental bounds \cite{pdg}: \be
\label{pdgBOUND}{\rm Br}(\mu \to e \gamma)<1.2 \times 10^{-11}\ \
{\rm and} \ \ {\rm Br}(\mu^- \to e^-e^+e^-)<1.0 \times 10^{-12}\ .
\ee Notice that the bound on $\mu \to eee$  is even stronger than
the famous bound on $\mu \to e \gamma$. In the context of models
like R-parity conserving MSSM in which the new particles can
appear only in even numbers in each vertex, these processes can
only take place  at the loop level. Since $\mu \to eee$ is a three
body decay, in such a model, ${\rm Br}(\mu \to eee)$ is suppressed
relative to ${\rm Br}(\mu \to e \gamma)$ by a factor typically of
order of $e^2/(16 \pi^2) \log (m_\mu/m_e)$  \cite{tobe}. However,
in the models that new particles can appear in odd numbers at each
vertex ({\it e.g.,} in R-parity violating MSSM) the process $\mu
\to eee$ can take place at tree level and as a result, its rate
can even exceed that of $\mu\to e\gamma$ \cite{andre}.

It is rather well-known that by measuring the angular distribution
of the final particles relative to the spin of the initial muon in
$\mu\to e \gamma$ and $\mu \to eee$, one can derive information on
the chiral nature of the effective Lagrangian leading to this
process \cite{andre,veto,review}. In the case of $\mu \to eee$, as
shown in the literature \cite{andre,zee,okada}, the angular
distribution of the final particles relative to the spin of the
initial muon also yields information on certain combinations of
the CP-violating phases.

 Recently, it has been shown in
\cite{khodam,khodemoon,Sacha} that if we  measure the polarization
of the emitted particles in $\mu \to e \gamma$ and $\mu N \to e
N$, we can derive information on the CP-violating parameters of
the theory.
 It was pointed out in \cite{khodam,khodemoon} that by
 measuring the spin of the more energetic final positron in
 $\mu^+\to e^+e^-e^+$, some information on the CP-violating phases
 can be derived. The analysis was performed in the framework of the
 models such as $R$-parity conserving MSSM, in which the dominant contribution to $\mu \to eee$ comes
 from a penguin diagram (i.e., $\mu^+\to \gamma^* e^+ \to
 e^+e^-e^+$).
 In this letter, we focus on the case that $
\mu^+ \to e^+e^- e^+$ happens at the tree level through the
exchange of heavy particles. We show that  the transverse
 polarizations of the emitted particles in $\mu^+ \to e^+
e^- e^+$ provide us with  information  on  the CP-violating
properties of the effective Lagrangian. The information derived
from the spins of the emitted electrons and positrons are
complementary to each other. In this letter, we focus on $\mu^+\to
e^+e^-e^+$. Similar arguments hold for $\mu^-\to e^-e^+e^-$.

 We show that the method
 proposed here is sensitive to a combination of the phases in the effective
Lagrangian which is different from those that can be derived by
methods discussed in the literature \cite{andre,zee,okada}. The
effectiveness of each method depends on the relative magnitude of
the different terms in the effective Lagrangian which in turn
depends on the details of the underlying model.

\section{The rate of $\mu^+\to e^+e^-e^+$}
Consider a general beyond SM scenario leading to $\mu^+ \to e^+e^-
e^+$. After integrating out the heavy states, the effect can be
described by an  effective Lagrangian of form \be {\cal L}={\cal
L}_{{\rm 1}} +{\cal L}_{{\rm 2}}\ee where
   \ba
\label{effectiveLag}{\cal L}_{\rm 1}&=& B_1 (\bar{\mu}{1+\gamma_5
\over 2}e)( \bar{e} {1-\gamma_5 \over 2}e)+B_2(
\bar{\mu}{1-\gamma_5 \over 2}e )(\bar{e} {1+\gamma_5 \over 2}e) +
\ \ \ \cr &\ &
 C_1
(\bar{\mu}{1+\gamma_5 \over 2}e)( \bar{e} {1+\gamma_5 \over
2}e)+C_2 (\bar{\mu}{1-\gamma_5 \over 2}e)( \bar{e} {1-\gamma_5
\over 2}e)  \cr ~ &+& G_1 (\bar{\mu}\gamma^\nu{1+\gamma_5 \over 2}
e)(\bar{e}\gamma_\nu{1+\gamma_5 \over 2} e)+G_2
(\bar{\mu}\gamma^\nu{1-\gamma_5 \over 2}
e)(\bar{e}\gamma_\nu{1-\gamma_5 \over 2} e)+ {\rm H.c.} \ea and
\be {\cal L}_{{\rm 2}}= A_L \bar{\mu}[ \gamma^{\mu},\gamma^{ \nu}]
{1+\gamma_5 \over 2} e F_{\mu \nu}+A_R \bar{\mu}
[\gamma^{\mu},\gamma^\nu] {1-\gamma_5 \over 2} e F_{\mu \nu}+{\rm
H.c.} \  \ee Notice that by using the identities
$(\sigma^\mu)_{\alpha \beta}(\sigma_\mu)_{\gamma \delta}\equiv 2
\epsilon_{\alpha \gamma} \epsilon_{\beta \delta}$ and $
(\bar{\sigma}^\mu)_{\alpha \beta}=\epsilon_{\beta \delta}
(\sigma^\mu)_{\delta \gamma} \epsilon_{\gamma \alpha}$  (where
$\epsilon_{11}=\epsilon_{00}=0$ and
$\epsilon_{10}=-\epsilon_{01}=1$) and employing the fact that the
fermions anti-commute, we can rewrite the terms on the first line
of Eq.~(\ref{effectiveLag}) as
$$-{B_1 \over 2}(\bar{\mu} {1+\gamma_5 \over  2} \gamma^\nu e)(
\bar{e} {1-\gamma_5 \over  2} \gamma_\nu e) -{B_2 \over
2}(\bar{\mu} {1-\gamma_5 \over  2} \gamma^\nu e)( \bar{e}
{1+\gamma_5 \over  2} \gamma_\nu e) \ .$$ In the literature, it
has been shown that by studying the angular distribution of the
final particles relative to the spin of the initial muon, one can
derive information on  ${\rm Re}[A_R B_1^*-A_L B_2^*]$, ${\rm
Re}[A_R G_2^*-A_L G_1^*]$,  ${\rm Im}[A_R B_1^*+A_LB_2^*]$ and
${\rm Im}[A_R G_2^*+A_LG_1^*]$ (see {\it e.g.,}
\cite{andre,zee,okada}). However, by this method the phases of
$C_1$ and $C_2$ cannot be derived. Moreover, if $A_L$ and $A_R$
are zero or suppressed, this method loses its effectiveness.

By studying the energy spectrum of the emitted particles, it will
be possible to differentiate between the different terms in ${\cal
L}_{\rm 1}$ and  ${\cal L}_{\rm 2}$. For example,  the $A_L$ and
$A_R$ couplings lead to a sharp peak in the distribution of the
square of the invariant mass of a $e^-e^+$ pair [{\it i.e.,} in
the distribution of $(P_{e^-}-P_{e^+})^2$] at $(m_\mu/2)^2$. Such
a peak does not appear if the dominant contribution comes from
Eq.~(\ref{effectiveLag}).
 The $A_L$ and $A_R$ couplings are generally loop suppressed but, as we see below,
 terms in   ${\cal
L}_{\rm 1}$ can appear in a wide range of models at tree level. In
the present letter, we only consider terms in ${\cal L}_{\rm 1}$.

 The $B_i$ and $C_i$ couplings  in ${\cal L}_{\rm 1}$ can
originate from the exchange of a heavy neutral scalar field (or
fields). Let us demonstrate this through a simple toy model.
Consider two heavy {\it complex} fields, $\phi_1$ and $\phi_2$
with  the following couplings
$$g_{\mu L} \phi_1 \bar{\mu} \frac{1+\gamma_5}{2} e+g_{\mu R}
\phi_2 \bar{\mu}\frac{1-\gamma_5}{2} e+ g_{e L} \phi_1 \bar{e}
\frac{1+\gamma_5}{2} e+g_{e R} \phi_2 \bar{e}\frac{1-\gamma_5}{2}
e +{\rm H.c.}$$ It is straightforward to show that the effective
Lagrangian resulting from integrating out the heavy states
$\phi_1$ and $\phi_2$ is of form (\ref{effectiveLag}) with
$$ B_1={g_{\mu L}g_{e L}^* \over m_{\phi_1}^2}, \ \ B_2={g_{\mu R}g_{e R}^* \over
m_{\phi_2}^2}, \ \ C_1=C_2=0 \ .$$ If we swap $\phi_1$ and
$\phi_2$ in the third and fourth terms ({\it i.e.,} taking ${\cal
L}=[g_{\mu L} \phi_1 \bar{\mu} ({1+\gamma_5}) e+g_{\mu R} \phi_2
\bar{\mu}({1-\gamma_5}) e+ g_{e L} \phi_2 \bar{e} ({1+\gamma_5})
e+g_{e R} \phi_1 \bar{e}({1-\gamma_5}) e +{\rm H.c.}]/2$), we find
$ C_1={g_{\mu L}g_{e R}^* / m_{\phi_1}^2}, \ C_2={g_{\mu R}g_{e
L}^* / m_{\phi_2}^2} \ {\rm and} \
 B_1=B_2=0.$ Taking $\phi_1$ and $\phi_2$ {\it real}, we find that
  $B_1$, $B_2$, $C_1$ and $C_2$ are all nonzero.
It can  similarly be shown that  $G_1$ and $G_2$ can originate
from the exchange of a doubly charged scalar field.
 The R-parity
violating MSSM at the tree level leads to couplings of form $B_1$
and $B_2$ \cite{andre}. That is while in this model, the $A_L$ and
$A_R$ couplings are loop suppressed.

 Notice that while the $B_i$
couplings  are chirality conserving, the $C_i$ and $G_i$ couplings
 are chirality-flipping.
Notice that under the parity transformation,  $B_i$ and $C_i$
transform as $B_1 \Leftrightarrow B_2 $, $G_1 \Leftrightarrow G_2
$ and $C_1 \Leftrightarrow C_2$. Thus, $|B_1-B_2|$, $|G_1-G_2|$
and $|C_1-C_2|$ can be considered as measures of the parity
violation. Under the CP transformation
$$ B_1\Rightarrow \eta B_1^*\ , \ \ B_2\Rightarrow \eta B_2^* \ ,
\ \ G_1\Rightarrow \eta G_1^*\ , \ \ G_2\Rightarrow \eta G_2^* \ ,
\ \  C_1\Rightarrow \eta C_1^* \ {\rm and} \ C_2\Rightarrow\eta
C_2^* \ , $$ where $\eta$ is a pure phase that comes from the
freedom in the definition of the CP-conjugate of the electron and
the muon. By rephasing the electron  and/or the muon field, either
of these couplings can be made real.  Thus, the Lagrangian in
Eq.~(\ref{effectiveLag}) contains five physical CP-violating
phases.

 \begin{figure}[ht]\hskip -5 cm
\begin{center}
 \centerline{\includegraphics[height= 5 cm,bb=60 70 305 312,clip=true]{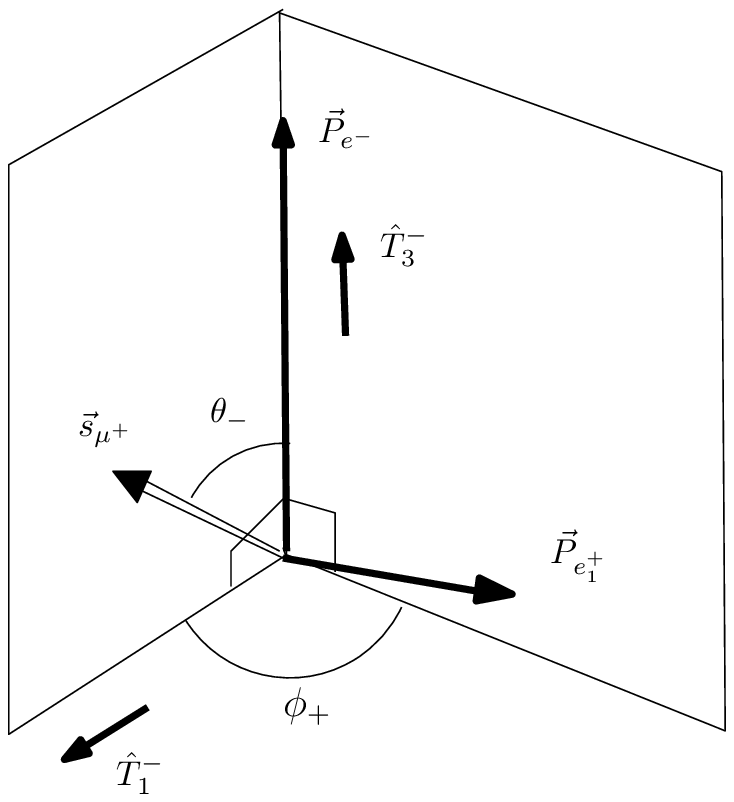}\hspace{5mm}
 \includegraphics[height= 5 cm,bb=60 70 305 312,clip=true]{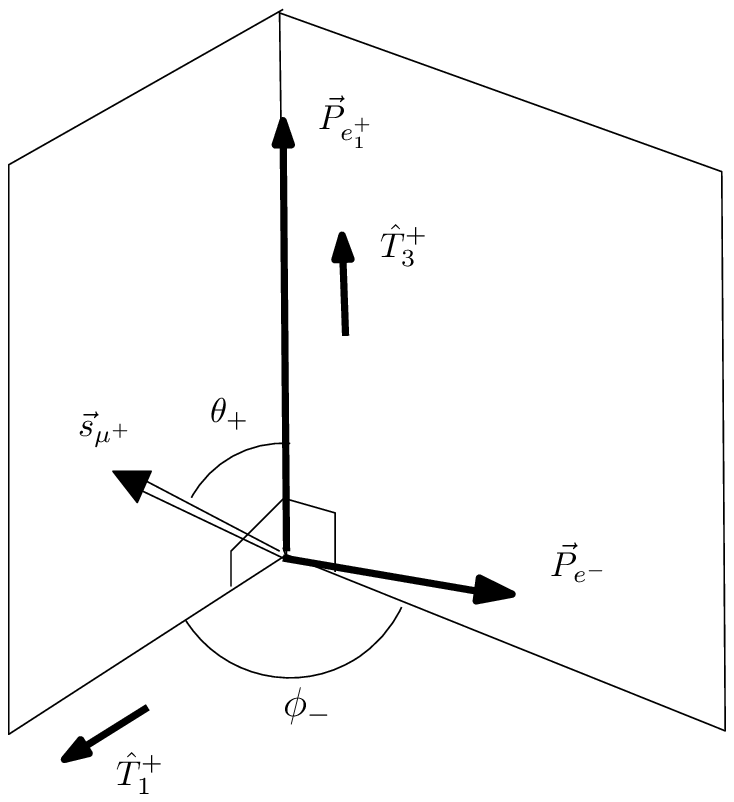}}
\centerline{\hspace{1.2cm}(a)\hspace{7cm}(b)}
   \caption{These figures schematically depict the direction of the  momenta of the final particles in the LFV decay
  $\mu^+\to e^+_1 e^- e^+_2$ relative to the
  spin of the anti-muon in its rest frame. Both figures
  correspond to a single decay a) illustrating
  $\hat{T}^-$, $\theta_-$ and $\phi_+$ and  b) illustrating
  $\hat{T}^+$, $\theta_+$ and $\phi_-$. } \label{3D}
  \end{center}
\end{figure}

From the formulas derived in the appendix, we find that the
differential decay rate of $\mu^+\to e^+e^-e^+$  is \ba
\label{nospin}&~&
\int_0^{2\pi}\sum_{\vec{s}_{e^-},\vec{s}_{e^+_1}\vec{s}_{e^+_2}}{d^3
\Gamma(\mu^+\to e^+_1e^-e^+_2) \over
dE_e~dE_{e^+_1}~d\Omega}d\phi_+ \cr
&=&\frac{1}{128\pi^4}\left[(|B_1|^2+|B_2|^2)(-m_\mu^3+4m_\mu^2E_{e^+_1}
+3m_\mu^2E_e-2E_e^2m_\mu-4m_\mu E_{e^+_1}^2-4m_\mu
E_eE_{e^+_1})\right.\cr &+&(|B_1|^2-|B_2|^2)\mathbb{P}_\mu
\cos\theta_-{(m_\mu^4-4m_\mu^3E_{e^+_1}-3m_\mu^3E_e+8m_\mu^2E_e
E_{e^+_1}+4m_\mu^2E_{e^+_1}^2+4m_\mu^2E_e^2-2m_\mu E_e^3-6m_\mu
E_e^2E_{e^+_1}) / E_e} \cr
&+&\left.\left[(|C_1|^2+16|G_2|^2)(1+\mathbb{P}_\mu \cos
\theta_-)+(|C_2|^2+16|G_1|^2)(1-\mathbb{P}_\mu \cos
\theta_-)\right]E_em_\mu(m_\mu-2E_{e})\right]\ea where $d\Omega$
is the differential solid angle determining the orientation of the
emitted electron. We have summed over the spins of the final
particles and integrated over $\phi_+$ which determines the
azimuthal angle of the emitted positrons (see Fig. \ref{3D}).

Integrating over the energies of the final particles, we find \ba
\sum_{\vec{s}_{e^-},\vec{s}_{e^+_1}, \vec{s}_{e^+_2}}
 {d \Gamma \over d\cos \Omega}&=&
 \frac{0.0208}{128 \pi^4}m_\mu^5\left[
(|B_1|^2+|B_2|^2+{|C_1|^2+|C_2|^2\over
2}+8|G_1|^2+8|G_2|^2)\right.\cr ~ &+&\left. \mathbb{P}_\mu\cos
\theta_-({|B_2|^2-|B_1|^2 \over 3}+{|C_1|^2-|C_2|^2\over
2}+8(|G_2|^2-|G_1|^2))\right]\ .\ea
 The total rate of $\int
(d\Gamma/d\Omega)d\Omega $ is given by
$(|B_1|^2+|B_2|^2+(|C_1|^2+|C_2|^2)/2)+8(|G_1|^2+|G_2|^2)$. From
the bound on ${\rm Br}(\mu\to eee)$ ({\it see}
Eq.~\ref{pdgBOUND}), we find
$$ |B_1|^2+|B_2|^2+{|C_1|^2+|C_2|^2\over 2}+8(|G_1|^2+|G_2|^2)<\frac{1}{(200~{\rm TeV})^4} \
.$$

From Eq.~(\ref{nospin}), we observe that the dependence of the
coefficients of $|B_1|^2+|B_2|^2$ and
$|C_1|^2+|C_2|^2+16(|G_1|^2+|G_2|^2)$ on $E_{e^+_1}$ and $E_e$ are
different. As a result, by studying the energy spectrum of the
final particles, it will be possible to extract $|B_1|^2+|B_2|^2$
and $|C_1|^2+|C_2|^2+16(|G_1|^2+|G_2|^2)$. If in addition to the
energy spectrum, the angular distribution of the electron relative
to $\vec{s}_\mu$ is measured, it will be possible to extract the
parity violating combinations $|B_1|^2-|B_2|^2$ and
$|C_1|^2-|C_2|^2+16(|G_2|^2-|G_1|^2)$. Of course the larger the
polarization of the initial muon, the higher the sensitivity of
the angular distribution to these combinations. Thus, in principle
by measuring the spectrum of the emitted particles and angular
distribution of the final electron, it will be possible to derive
the absolute values of the couplings. The angular distribution of
the final positrons also give information on $|B_1|^2-|B_2|^2$ and
$|C_1|^2-|C_2|^2+16(|G_2|^2-|G_1|^2)$ (see the appendix). Thus,
the following combinations can be measured from the study of the
angular distribution plus energy spectrum measurements: \be
|B_1|^2, \ |B_2|^2, \ |C_1|^2+16|G_2|^2\ {\rm and} \
|C_2|^2+16|G_1|^2\ . \label{measureable} \ee

 However,
without sensitivity to the spins of the final particles, it is not
possible to derive information on the phases of the couplings.
 In sect.~\ref{s-e}, we explore what can be learned from
the polarization of the emitted electron. In sect.~\ref{s+e}, we
discuss the polarization of the positron.

\subsection{Polarization of the emitted electron \label{s-e}}
 Consider the
decay $\mu^+ \to e^+ e^-e^+$ in the rest frame of the muon, where
the final electron makes an angle of $\theta_-$ with the spin of
the initial muon. Let us take the $\hat{T}_{3}^-$ direction
parallel to the momentum of the emitted electron and
$\hat{T}_{2}^-\equiv\hat{T}_{3}^-\times
\vec{s}_\mu/|\hat{T}_{3}^-\times \vec{s}_\mu|$ (see
Fig.~\ref{3D}). The spin of the electron is determined by
$(c_{e^-} \ d_{e^-})$ with $(|c_{e^-}|^2+|d_{e^-}|^2)^{1/2}=1$.
 That is the  components of the
spin of the electron are
 \be\hat{T}_{3}^-\cdot \vec{s}_{e^-}=|c_{e^-}|^2-|d_{e^-}|^2, \
 \hat{T}_{1}^-\cdot \vec{s}_{e^-}=2{\rm Re}[c_{e^-}^* d_{e^-}] \ {\rm and} \ \hat{T}_{2}^-\cdot \vec{s}_{e^-}
 =2{\rm Im}[c_{e^-}^* d_{e^-}].\label{spin-electron} \ee

  From the formulas in the appendix, we find that the differential
   decay rate in the rest frame of the muon is \ba \label{diffGamma4} &~&
  \sum_{\vec{s}_{e^+_1}, \vec{s}_{e^+_2}}  {d
\Gamma(\mu^+\to e^+_1 e^- e^+_2) \over d\cos \Omega}=
\int_0^{2\pi} \int_0^{m_\mu/2} \int_{m_\mu/2 -E_e}^{m_\mu/2}
\sum_{\vec{s_{e^+_1}}, \vec{s_{e^+_2}}} |M|^2 E_{e_1^+} E_e
(m_\mu-E_{e_1^+}-E_e)dE_{e_1^+} dE_e d\phi_+ \cr &=&
\frac{m_\mu^5}{128\pi^4} \left[
(0.0208)[|B_1|^2|c_{e^-}|^2+|B_2|^2 |d_{e^-}|^2 +
(|C_1|^2/2+8|G_2|^2)|d_{e^-}|^2+ (|C_2|^2/2+8 |G_1|^2)|c_{e^-}|^2]
\right. \cr &+& (0.0208)\mathbb{P}_\mu \cos\theta_- [
{|B_2|^2|d_{e^-}|^2/ 3} -{|B_1|^2|c_{e^-}|^2/
3}+(|C_1|^2/2+8|G_2|^2)|d_{e^-}|^2-(|C_2|^2/2+8
|G_1|^2)|c_{e^-}|^2] \cr &-& (0.0832)\mathbb{P}_\mu \sin \theta_-
({\rm Re}[G_1 C_1^*d_{e^-} c_{e^-}^*]+{\rm Re}[G_2 C_2^*d_{e^-}^*
c_{e^-}])\cr&+& \left. 0.055{\rm Re}[B_1 B_2^*d_{e^-}
c_{e^-}^*]\mathbb{P}_\mu \sin \theta_- \right]\ ,\ea where
$\mathbb{P}_\mu$ is the polarization of the initial muon and
 $d\Omega$ represents the differential solid angle of the momentum of the electron relative
 to the spin of the muon. To obtain this equation, we have summed over the spins of the
emitted positrons and integrated over the azimuthal angle that the
plane containing the momenta of these positrons makes with the
plane of the spin of the anti-muon and the momentum of the
electron.  See the appendix for the details. Notice that there is
no interference term between the chirality-flipping and
chirality-conserving couplings.

From Eq.~(\ref{16}) in the appendix, we find that the longitudinal
polarization of the electron is  \be \langle s_{T_3^-} \rangle
\equiv{{d \Gamma}/{d \Omega}|_{\stackrel{c_{e^-}=1}{d_{e^-}=0}}-
{d \Gamma}/{d \Omega}|_{\stackrel{c_{e^-}=0}{d_{e^-}=1}}\over {d
\Gamma}/{d \Omega}|_{\stackrel{c_{e^-}=1}{d_{e^-}=0}}+ {d
\Gamma}/{d \Omega}|_{\stackrel{c_{e^-}=0}{d_{e^-}=1}} }= {
P_1^--\mathbb{P}_\mu\cos \theta_- P_2^-\over P_3^-
+\mathbb{P}_\mu\cos \theta_- P_4^-} \ ,\label{T-3} \ee where \ba
P_1^- &=&\ |B_1|^2  - |B_2|^2 -
\frac{|C_1|^2}{2}+\frac{|C_2|^2}{2} + 8|G_1|^2 - 8|G_2|^2\cr
 P_2^-
&=&\ \frac{|B_1|^2}{3}+
\frac{|B_2|^2}{3}+\frac{|C_1|^2}{2}+\frac{|C_2|^2}{2}+
8|G_1|^2+8|G_2|^2 \cr P_3^- &=& \ |B_1|^2 + |B_2|^2 +
\frac{|C_1|^2}{2} + \frac{|C_2|^2}{2} + 8|G_1|^2+ 8|G_2|^2\cr
P_4^- &=& -\frac{|B_1|^2}{3} +
\frac{|B_2|^2}{3}+\frac{|C_1|^2}{2}- \frac{|C_2|^2}{2}-
8|G_1|^2+8|G_2|^2\ . \ea
 $\langle s_{T_3^-}\rangle$ is sensitive only to the absolute
values of the couplings. Notice that by measuring $\langle
s_{T_3^-}\rangle$ and its angular distribution, one can derive the
same combinations that are listed in Eq.~(\ref{measureable}) {\it
i.e.,} the combinations that can be extracted from the angular
distribution and energy spectrum (without measuring the spin of
the final particles).
 Derivation of these combinations from
the measurement of $\langle s_{T^-_3} \rangle$  can be considered
as a cross-check for the derivation  from the energy spectrum.
Notice that even in the  $\mathbb{P}_\mu=0$ limit, $\langle
s_{T_3^-}\rangle$ is still non-vanishing and provides us with
information on the parity violating combination $P_1$. That is
while in this limit,
 the angular
distribution of the electron is uniform and has no sensitivity to
parity violation in the effective Lagrangian (\ref{effectiveLag})
(see Eq.~(\ref{nospin})).

 The transverse polarizations are
\ba \langle s_{T^-_1} \rangle &\equiv& {{d \Gamma}/{d
\Omega}|_{\stackrel{c_{e^-}=1/\sqrt{2}}{d_{e^-}=1/\sqrt{2}}}- {d
\Gamma}/{d
\Omega}|_{\stackrel{c_{e^-}=1/\sqrt{2}}{d_{e^-}=-1/\sqrt{2}}}\over
{d \Gamma}/{d
\Omega}|_{\stackrel{c_{e^-}=1/\sqrt{2}}{d_{e^-}=1/\sqrt{2}}}+ {d
\Gamma}/{d
\Omega}|_{\stackrel{c_{e^-}=1/\sqrt{2}}{d_{e^-}=-1/\sqrt{2}}} }
\cr ~ &=&   \label{T-1}{{\rm Re}\left[(2.6)B_1^*B_2-4 G_1^*
C_1-4G_2 C_2^*\right]\sin \theta_-\mathbb{P}_\mu\over
P_3^-+P_4^-\mathbb{P}_\mu\cos \theta_- }\ \ea and
 \ba \langle s_{T^-_2} \rangle &\equiv& {{d
\Gamma}/{d
\Omega}|_{\stackrel{c_{e^-}=1/\sqrt{2}}{d_{e^-}=i/\sqrt{2}}}- {d
\Gamma}/{d
\Omega}|_{\stackrel{c_{e^-}=1/\sqrt{2}}{d_{e^-}=-i/\sqrt{2}}}\over
{d \Gamma}/{d
\Omega}|_{\stackrel{c_{e^-}=1/\sqrt{2}}{d_{e^-}=i/\sqrt{2}}}+ {d
\Gamma}/{d
\Omega}|_{\stackrel{c_{e^-}=1/\sqrt{2}}{d_{e^-}=-i/\sqrt{2}}} }\cr
~ &=&  \label{T-2}{{\rm Im}\left[(2.6)B_1^*B_2-4  G_1^*C_1-4
G_2C_2^*]\right]\sin \theta_-\mathbb{P}_\mu\over
P_3^-+P_4^-\mathbb{P}_\mu\cos \theta_- }\ .\ea  The transverse
polarizations are maximal for electrons emitted in the direction
perpendicular to the spin of the muon: {\it i.e.,
$\theta_-=\pi/2$.} Notice that
 $\langle s_{T^-_{1}} \rangle$  and  $\langle s_{T^-_{2}} \rangle$
 provide independent information on the
real and imaginary parts of the same combination; {\it i.e.,}
$(2.6 B_1^*B_2-4  G_1^*C_1-4 G_2C_2^*)$. Notice that both the
moduli and the phase of this combination contain extra information
beyond the combinations listed in Eq.~(\ref{measureable}). In the
context of various models $G_i$ and $C_i$ can vanish. For example,
as explained in the previous section, if the effective Lagrangian
comes from integrating out  neutral scalars, the $G_i$ couplings
will vanish. Within such models,  $\langle s_{T_1^-}\rangle
\propto {\rm Re}[B_1 B_2^*]$ and  $\langle s_{T_2^-}\rangle
\propto {\rm Im}[B_1 B_2^*]$. Considering that $|B_i|$ can be
independently measured,  the  derivation of $\arg[B_1 B_2^*]$ from
both $\langle s_{T^-_{1}} \rangle$ and $\langle s_{T^-_{2}}
\rangle$ can be considered as a cross-check.

In the limit of unpolarized muon ({\it i.e.,} $\mathbb{P}_\mu\to
0$), $\langle s_{T^-_1}\rangle$ and $\langle s_{T^-_2}\rangle$
vanish. This is understandable because in the $\mathbb{P}_\mu=0$
limit, there is no preferred directions so we cannot define
$\hat{T}_{1}^-$ and $\hat{T}_{2}^-$ (see Fig. \ref{3D}). Thus, to
derive the CP-violating phase, $\arg[B_1B_2^*]$, it is necessary
to have a source of polarized muon which is quite feasible. For
example if the muons are produced by the decay of pions at rest,
they will be almost 100 \% polarized. In fact, this is the case
for the on-going MEG experiment.

By measuring the polarization of the emitted electron,  it is not
possible to derive the values of all the
 CP-violating phases   of the  Lagrangian
(\ref{effectiveLag}). Measurement of the angular distribution of
the positrons does not provide any further information. As we
shall see next, the transverse polarizations of the emitted
positrons provide complementary information on the phases.


\subsection{Polarization of the positron \label{s+e}}
 Consider the
decay $\mu^+ \to e^+_1 e^-e^+_2$ in the rest frame of the muon
where one of the final positrons makes an angle of $\theta_+$ with
the spin of the initial muon. Let us take the $\hat{T}_{3}^+$
direction parallel to the momentum of $e^+_1$ and
$\hat{T}_{2}^+\equiv\hat{T}_{3}^+\times
\vec{s}_\mu/|\hat{T}_{3}^+\times \vec{s}_\mu|$ (see
Fig.~\ref{3D}). The spin of $e^+_1$ is determined by $(c_{e^+_1} \
d_{e^+_1})$ with $(|c_{e^+_1}|^2+|d_{e^+_1}|^2)^{1/2}=1$.
 The  components of the spin are
 \be \label{spin-positron}\hat{T}_{3}^+\cdot \vec{s}_{e^+_1}=|c_{e^+_1}|^2-|d_{e^+_1}|^2,
 \ \hat{T}_{1}^+\cdot \vec{s}_{e^+_1}=2{\rm Re}[c_{e^+_1}^*
 d_{e^+_1}]\
{\rm and} \ \hat{T}_{2}^+\cdot \vec{s}_{e^+_1}
 =2{\rm Im}[c_{e^+_1}^* d_{e^+_1}]\ . \ee
 From the formula in the appendix, we find that the differential
 decay rate in the rest frame of the muon is
\ba
  \sum_{\vec{s}_{e^-}, \vec{s}_{e^+_2}}  {d
\Gamma(\mu^+\to e^+_1 e^- e^+_2) \over d\cos \Omega} &=&
\int_0^{2\pi} \int_0^{m_\mu/2} \int_{m_\mu/2 -E_e}^{m_\mu/2}
\sum_{\vec{s}_{e^-}, \vec{s}_{e^+_2}} |M|^2 E_{e_1^+} E_e
(m_\mu-E_{e_1^+}-E_e)dE_{e} dE_{e^+_1} d\phi_- \cr ~ &=&
\label{diffGamma} (0.0104){m_\mu^5\over
128\pi^4}\left[(|C_1|^2|c_{e^+_1}|^2+|C_2|^2|d_{e^+_1}|^2)-
(|C_1|^2|c_{e^+_1}|^2-|C_2|^2|d_{e^+_1}|^2){\mathbb{P}_\mu
\cos\theta_+ \over 3}\right.\cr ~ &+&
|B_1|^2+|B_2|^2+[|B_1|^2(|c_{e^+_1}|^2-{|d_{e^+_1}|^2/
3})-|B_2|^2(|d_{e^+_1}|^2-{|c_{e^+_1}|^2/ 3})]\mathbb{P}_\mu
\cos\theta_+ \cr ~ &+&
16(|G_2|^2|c_{e^+_1}|^2+|G_1|^2|d_{e^+_1}|^2)\cr ~ &-&
\frac{16}{3}(|G_2|^2|c_{e^+_1}|^2-|G_1|^2|d_{e^+_1}|^2)\mathbb{P}_\mu
\cos\theta_+ \cr ~ &+&  ({\rm Re}[B_1C_2^*c_{e^+_1}d_{e^+_1}^*]+
{\rm Re}[B_2C_1^*c_{e^+_1}^*d_{e^+_1}])\mathbb{P}_\mu \sin\theta_+
\cr ~ &+& \left. 24({\rm Re}[G_1 B_1^*c_{e^+_1}^*d_{e^+_1}]-{\rm
Re}[G_2 B_2^*c_{e^+_1}d_{e^+_1}^*] )\mathbb{P}_\mu
\sin\theta_+\right] \ea where $d\Omega$ represents the
differential solid angle of the momentum of $e^+_1$ relative
 to the spin of the muon.

From the above equation, we find that the longitudinal
polarization of $e^+_1$ is\be \label{T+3}\langle s_{T_3^+}\rangle
={{d \Gamma}/{d \Omega}|_{\stackrel{c_{e^+_1}=1}{d_{e^+_1}=0}}- {d
\Gamma}/{d \Omega}|_{\stackrel{c_{e^+_1}=0}{d_{e^+_1}=1}}\over {d
\Gamma}/{d \Omega}|_{\stackrel{c_{e^+_1}=1}{d_{e^+_1}=0}}+ {d
\Gamma}/{d \Omega}|_{\stackrel{c_{e^+_1}=0}{d_{e^+_1}=1}}
}={P_1^++P_2^+\mathbb{P}_\mu\cos\theta_+/3\over P_3^++
P_4^+\mathbb{P}_\mu\cos\theta_+/3} \ ,\ee where \ba P_1^+
&=&|C_1|^2-|C_2|^2+16|G_2|^2-16|G_1|^2~~~~~ \cr
P_2^+&=&-|C_2|^2-|C_1|^2-16|G_2|^2-16|G_1|^2+4|B_1|^2+4|B_2|^2 \cr
P_3^+&=&|C_1|^2+|C_2|^2+16(|G_1|^2+|G_2|^2)+2(|B_1|^2+|B_2|^2) \cr
P_4^+&=&|C_2|^2-|C_1|^2-16|G_2|^2+16|G_1|^2+2|B_1|^2-2|B_2|^2 \
.\ea
The transverse polarizations are \ba  \label{T+1}\langle
s_{T_1^+}\rangle &=& {{d \Gamma}/{d
\Omega}|_{\stackrel{c_{e^+_1}=1/\sqrt{2}}{d_{e^+_1}=1/\sqrt{2}}}-
{d \Gamma}/{d
\Omega}|_{\stackrel{c_{e^+_1}=1/\sqrt{2}}{d_{e^-}=-1/\sqrt{2}}}\over
{d \Gamma}/{d
\Omega}|_{\stackrel{c_{e^+_1}=1/\sqrt{2}}{d_{e^+_1}=1/\sqrt{2}}}+
{d \Gamma}/{d
\Omega}|_{\stackrel{c_{e^+_1}=1/\sqrt{2}}{d_{e^+_1}=-1/\sqrt{2}}}}
\cr &=&{{\rm
Re}\left[B_1C_2^*+B_2^*C_1+24G_1^*B_1-24G_2B_2^*\right]
)\sin\theta_+\mathbb{P}_\mu \over P_3^++
P_4^+\mathbb{P}_\mu\cos\theta_+/3} \ea
and
\ba \label{T+2}\langle s_{T^+_2}\rangle &=& {{d \Gamma}/{d
\Omega}|_{\stackrel{c_{e^+_1}=1/\sqrt{2}}{d_{e^+_1}=i/\sqrt{2}}}-
{d \Gamma}/{d
\Omega}|_{\stackrel{c_{e^+_1}=1/\sqrt{2}}{d_{e^+_1}=-i/\sqrt{2}}}\over
{d \Gamma}/{d
\Omega}|_{\stackrel{c_{e^+_1}=1/\sqrt{2}}{d_{e^+_1}=i/\sqrt{2}}}+
{d \Gamma}/{d
\Omega}|_{\stackrel{c_{e^+_1}=1/\sqrt{2}}{d_{e^+_1}=-i/\sqrt{2}}}}
\cr ~ &=&{{\rm Im}\left[B_1C_2^*+B_2^*C_1 +24G_1^*B_1-24
G_2B_2^*\right]\sin\theta_+\mathbb{P}_\mu\over P_3^++
P_4^+\mathbb{P}_\mu\cos\theta_+/3}\ .\ea Like the case of the
electron discussed in sect. \ref{s-e}, the longitudinal
polarization, $\langle s_{T^+_3}\rangle$, does not give
information on the CP-violating phases and can be used only as a
cross-check for the derivation of the combinations listed in
Eq.~(\ref{measureable}) by the methods discussed earlier. The
ratio of $\langle s_{T^+_1}\rangle$ and $\langle s_{T_2^+}\rangle$
gives
 $\arg\left[B_1C_2^*+B_2^*C_1 +24G_1^*B_1-24
G_2B_2^*\right]$. Considering that the absolute value of this
combination is also unknown, $|\langle
s_{T_2^+}\rangle|^2+|\langle s_{T_1^+}\rangle|^2$ provides an
independent piece of information.
 We have integrated over $\phi_-$
which means the measurement of the direction of the emitted
electron is not necessary for this analysis.

\section{Conclusions and Prospects}
A large  variety of the beyond standard models predict a sizeable
rate for $\mu^+\to e^+e^-e^+$ exceeding the present experimental
bound. In principle, by studying the energy spectrum of the final
particles and their angular distribution, it is possible to derive
the form of the terms in the effective Lagrangian leading to this
process and extract information on the {\it absolute values} of
the  couplings (see Eq.~(\ref{measureable})). The effective
Lagrangian responsible for $\mu^+\to e^+e^-e^+$ can include  new
CP-violating phases.
In order to derive information on  the CP-violating phases, we
have suggested to measure the polarization of the emitted
particles. In this letter, we have focused on the effective
Lagrangian in Eq.~(\ref{effectiveLag}) that can result from
integrating out heavy scalar fields with LFV couplings at the tree
level.  The rest of the terms  ({\it i.e.,} $A_L$ and $A_R$) are
expected to be loop suppressed and are neglected in this analysis.
We have shown that the transverse polarization of the emitted
electron in $\mu^+\to e^+e^-e^+$ is sensitive to
$\arg[2.6B_1B_2^*-4 G_1^*C_1-4 G_2C_2^*]$ [see
Eqs.~(\ref{T-1},\ref{T-2})]. That is while the transverse
polarizations of the emitted positron is given by
$\arg[B_1C_2^*+B_2^* C_1 +24 G_1^*B_1-24 G_2 B_2^*]$.
 From Eqs.
(\ref{T-1},\ref{T-2},\ref{T+1},\ref{T+2}), we observe that if the
initial muon is unpolarized ({\it i.e.,} $\mathbb{P}_\mu=0$) the
transverse polarizations of the emitted particles vanish.
 Thus, in order to derive the CP-violating
phases, a source of polarized muons is required.

In sum, the effective Lagrangian in Eq.~(\ref{effectiveLag})
includes six new couplings and five physical phases. By measuring
the energy spectrum of the final particles and the angular
distributions relative to the spin of the initial muon, one can
derive the CP-conserving combinations listed in
Eq.~(\ref{measureable}): {\it i.e.,} four out of the six
CP-conserving quantities. Neglecting the loop suppressed $A_L$ and
$A_R$ couplings, the angular distribution cannot provide
information on the phases. The transverse polarizations of the
emitted particles provide four independent pieces of information
on the phases and couplings. This information is not enough to
reconstruct all the couplings but considerably reduces the
degeneracy in the parameter space.

 We have also
discussed the longitudinal polarization of the emitted particles.
The longitudinal polarizations do not depend on the phases of the
effective couplings. It is noteworthy that even in the
$\mathbb{P}_\mu=0$ limit, the longitudinal polarizations of the
positron, $\langle s_{T^+_3}\rangle$, and the electron, $\langle
s_{T^-_3}\rangle$, are nonzero and respectively yield information
on the parity violating combinations
$|C_1|^2-|C_2|^2+16|G_2|^2-16|G_1|^2$ and
$|C_1|^2-|C_2|^2+16|G_2|^2-16|G_1|^2+2(|B_2|^2-|B_1|^2)$. Remember
that in the $\mathbb{P}_\mu=0$ limit, there is no preferred
direction so the angular distribution of the final particles is
uniform and does not yield information on the parity-violating
combinations.

There are running and/or under construction experiments that aim
to probing signals for $\mu \to e \gamma$ \cite{MEGhomepage} and
$\mu-e$ conversion on nuclei \cite{Prism} several orders of
magnitudes below the present bounds on their rates. However, as
shown in \cite{andre}, it is possible that while $\mu^+\to e^+
e^-e^+$ is round the corner, the rates of $\mu \to e \gamma$ and
$\mu-e$ conversion are too low to be probed. In fact as shown in
\cite{andre}, the three experiments provide us with complementary
information on the parameters of the effective LFV Lagrangian. If
the muons are produced from the decay of pions at rest (like the
case of the running  MEG experiment \cite{MEGhomepage}), the
initial muons in $\mu \to eee$ will be polarized. On the other
hand, there are well-established techniques to measure the
polarization of the emitted particles in this energy range. In
fact, such polarimetry has been used to measure the Michel
parameters since 80's \cite{transeversemichel}. As a result, if
the rate of $\mu^+\to e^+e^-e^+$ is close to the present bound and
a hypothetical experiment finds statistically large number of such
a process, performing the analysis proposed in this paper sounds
possible.

In this letter, we have focused on the LFV three-body decay of the
anti-muon, $\mu^+\to e^+e^-e^+$. The same discussion applies to
the decay of the muon, $\mu^-\to e^-e^+e^-$. In this mode, the
transverse polarizations of the electrons would give
$\arg[B_1C_2^*+B_2^* C_1 +24 G_1^*B_1-24 G_2 B_2^*]$ and the
transverse polarizations of the emitted positrons would give
$\arg[2.6B_1B_2^*-4 G_1^*C_1-4 G_2C_2^*]$. The method of measuring
the polarization described in \cite{transeversemichel} is based on
studying the distribution of the photon pair from the annihilation
of the emitted positron on an electron in a target. If this method
is to be employed, only the polarization of the positron can be
measured. Thus, to derive both combinations, the experiment has to
run in both muon and anti-muon modes.

The three-body LFV decay modes of the tau lepton such as $\tau\to
e \bar{e} e$ or $\tau \to \mu \bar{\mu} \mu$ can also shed light
on the underlying theory. Recently it has been shown that by
studying the angular distribution of the final particles in $\tau
\to \mu \bar{\mu}\mu$ at the LHC, one can discriminate between
various models \cite{Ben}. Discussions in the present letter also
apply to the decay modes $\tau \to e \bar{e} e$ and $\tau \to \mu
\bar{\mu}\mu$.

\section*{Appendix}

In this appendix, we derive the decay rate of $\mu^+ \to e^+_1 e^-
e^+_2$. We first  derive the decay rate into an electron of
definite spin, summing over the spins of $e^+_1$ and $e^+_2$. We
then concentrate on the spin of $e^+_1$ and derive the decay rate
into a positron of definite spin, summing over the spins of the
electron and the other positron.

With  effective Lagrangian (\ref{effectiveLag}), we find \ba
M(\mu^+\to e^+_1e^- e^+_2)&=&B_1\bar{\mu}{1+\gamma_5\over 2}
e_1\bar{e}{1-\gamma_5 \over 2} e_2-B_1\bar{\mu}{1+\gamma_5\over 2}
e_2\bar{e}{1-\gamma_5 \over 2} e_1\cr
~~~~&+&B_2\bar{\mu}{1-\gamma_5\over 2} e_1\bar{e}{1+\gamma_5 \over
2} e_2-B_2\bar{\mu}{1-\gamma_5\over 2} e_2\bar{e}{1+\gamma_5 \over
2} e_1\cr ~~~~~&+& C_1\bar{\mu}{1+\gamma_5\over 2}
e_1\bar{e}{1+\gamma_5 \over 2} e_2-C_1\bar{\mu}{1+\gamma_5\over 2}
e_2\bar{e}{1+\gamma_5 \over 2} e_1\cr
~~~~&+&C_2\bar{\mu}{1-\gamma_5\over 2} e_1\bar{e}{1-\gamma_5 \over
2} e_2-C_2\bar{\mu}{1-\gamma_5\over 2} e_2\bar{e}{1-\gamma_5 \over
2} e_1 \cr ~~~~ &-&4 G_1 \bar{\mu}c {1-\gamma_5 \over 2} \gamma^0
e^* e_1^T c {1+\gamma_5 \over 2} e_2-4G_2 \bar{\mu}c {1+\gamma_5
\over 2} \gamma^0 e^* e_1^T c {1-\gamma_5 \over 2} e_2  . \ea

\subsection*{Decay rate into $e^-$ with a given spin}

Consider the decay $\mu^+\to e_1^+e^-e_2^+$ in the rest frame of
the muon.  Since we are interested in the spin of the electron, it
is convenient to use the coordinate system defined as
$\hat{T}^-_3\equiv \vec{P}_{e^-}/|\vec{P}_{e^-}|$,
$\hat{T}^-_2\equiv (\hat{T}_3^-\times
\vec{s}_\mu)/|\hat{T}_3^-\times \vec{s}_\mu|$ and
$\hat{T}_1^-\equiv \hat{T}_2^-\times \hat{T}_3^-$. In this
coordinate system, \ba P_{\mu^+}=(m_\mu,0,0,0)& \ \ \ &
P_{e_1^+}=E_{e_1^+}(1,\sin\alpha \cos \phi_+,\sin\alpha \sin
\phi_+,\cos \alpha)\ \ \ \ \ \cr P_{e^-}=(E_e,0,0,E_e)& \ \ \ &
P_{e_2^+}=(m_\mu-E_{e_1^+}-E_e,-E_{e_1^+}\sin\alpha \cos
\phi_+,-E_{e_1^+}\sin\alpha \sin \phi_+,-E_e-E_{e_1^+}\cos \alpha)
\ea where the electron mass is neglected (see Fig.~\ref{3D}-a).
Writing the kinematics and neglecting effects of
$O(m_e^2/m_\mu^2)\ll 1$, we find \be \label{alpha}\cos
\alpha={m_\mu^2-2m_\mu E_{e_1^+}-2m_\mu E_e +2 E_{e_1^+} E_e \over
2 E_{e_1^+} E_e}\ .\ee
  Summing over the spins
of the emitted positrons, we find that  \ba \! &~&
(2\pi)^4\int_0^{2\pi}(2E_{e^+_1})(2E_{e^+_2})
\sum_{\vec{s}_{e^+_1},\vec{s}_{e^+_2}}|M|^2d \phi_+\cr \! &=&
(|B_1|^2|c_{e^-}|^2+|B_2|^2|d_{e^-}|^2)E_{e^+_1}\left[
(m_\mu-E_{e^+_1}(1-\cos\alpha))+
(1-\cos\alpha)(m_\mu-E_e-E_{e^+_1})\right] \cr \! &+&
(|B_1|^2|c_{e^-}|^2-|B_2|^2|d_{e^-}|^2)E_{e^+_1}\left[ \cos \alpha
[m_\mu-E_{e^+_1}(1-\cos\alpha)]- (1-\cos\alpha)(E_{e^+_1}\cos
\alpha +E_e)\right] \mathbb{P}_\mu \cos \theta_- \cr\!
&+&\left[(|C_1|^2+16|G_2|^2)|d_{e^-}|^2(1+\mathbb{P}_\mu \cos
\theta_-)+(|C_2|^2+16|G_1|^2)|c_{e^-}|^2(1-\mathbb{P}_\mu \cos
\theta_-)\right] E_{e^+_1}[m_\mu -E_{e}(1-\cos \alpha)]~~~
 \cr
 \! &-& 2\mathbb{P}_\mu {\rm Re}[B_1
B_2^*d_{e^-} c_{e^-}^*]E_{e^+_1}(1-\cos \alpha)(E_{e^+_1}(1-\cos
\alpha)-m_\mu)\sin \theta_-
 \cr
 \! &+& 8\mathbb{P}_\mu ({\rm Re}([G_1
C_1^*d_{e^-} c_{e^-}^*]+{\rm Re}[G_2 C_2^*d_{e^-}^*
c_{e^-}])E_{e^+_1}(E_{e}(1-\cos \alpha)-m_\mu)\sin \theta_- \ ,
~~~~~~~~\label{MM}
 \ea
where  $(c_{e^-} \ d_{e^-})$ determines the spin of the emitted
electron [see Eq.~(\ref{spin-electron})].

 The differential  rate of the decay into an electron in
a direction that makes an angle of $\theta_-$ with the spin of the
initial muon is $$\sum_{\vec{s}_{e^+_1},
\vec{s}_{e^+_2}}\frac{d\Gamma(\mu^+\to
e^+_1e^-e^+_2)}{d\Omega}=\int_0^{2\pi}
\int_0^{m_\mu/2}\int_{m_\mu/2-E_e}^{m_\mu/2}
\sum_{\vec{s}_{e^+_1},\vec{s}_{e^+_2}} |M|^2 E_eE_{e^+_1}(m_\mu
-E_e -E_{e^+_1}) dE_{e^+_1}dE_e d\phi_+, $$ where  $d\Omega$ is
the differential solid angle determining the orientation of the
emitted electron. The factor $ E_eE_{e^+_1}(m_\mu -E_e
-E_{e^+_1})$ comes from the momentum-space volume for a three body
decay [{\it i.e.,} from $ \delta^4(P_\mu -P_e-P_{e^+_1}-P_{e^+_2})
d^3P_ed^3P_{e^+_1}d^3P_{e^+_2}$].
 Inserting $|M|^2$ from
Eq.~(\ref{MM}), we obtain \ba \label{16}&~& \sum_{\vec{s}_{e^+_1},
\vec{s}_{e^+_2}}{d\Gamma(\mu^+\to e^+_1e^-e^+_2)\over d\Omega}\cr
&=&{m_\mu^5\over
8(2\pi)^4}\left[(|c_{e^-}|^2|B_1|^2+|d_{e^-}|^2|B_2|^2)\int_0^{1/2}\int_{1/2-y}^{1/2}
(-1+3y+4x-2y^2-4x^2-4xy)dx~dy\right. \cr &+&\mathbb{P}_\mu\cos
\theta_-
(|B_1|^2|c_{e^-}|^2-|B_2|^2|d_{e^-}|^2)\int_0^{1/2}\int_{1/2-y}^{1/2}[1-2x-2y+2xy
-\frac{2x+2y-1}{y}(1-2x-2y+2xy+y^2)] dx~ dy\cr &+&
((|C_1|^2+16|G_2|^2)|d_{e^-}|^2(1+\mathbb{P}_\mu\cos\theta_-)+(|C_2|^2+16|G_1|^2)|c_{e^-}|^2
(1-\mathbb{P}_\mu\cos\theta_-) )
 \int_0^{1/2}\int_{1/2-y}^{1/2}y(1-2y) dx~dy\cr
\cr
 \! &-& 8\mathbb{P}_\mu ({\rm Re}[G_1
C_1^*d_{e^-} c_{e^-}^*]+{\rm Re}[G_2 C_2^*d_{e^-}^* c_{e^-}])\sin
\theta_-\int_0^{1/2}\int_{1/2-y}^{1/2}y(1-2y) dx~dy \cr
 &-&\left.2\mathbb{P}_\mu {\rm
Re}[B_1B_2^*d_{e^-}c_{e^-}^*] \sin \theta_-
\int_0^{1/2}\int_{1/2-y}^{1/2} (2x+2y-1)\frac{2x-1}{y}
dx~dy\right] \ . \cr \ea
 The integrals are all finite and lead to the numbers in Eq.~(\ref{diffGamma4})

\subsection*{Decay rate into $e^+_1$ with a given spin} Let us now
concentrate on one of the positrons, $e^+_1$, whose momentum makes
an angle of $\theta_+$ with $\vec{s}_\mu$ (see Fig. \ref{3D}-b).
To perform this analysis, it is convenient to work in the
following coordinate system:
$\hat{T}^+_3\equiv\vec{P}_{e^+_1}/|\vec{P}_{e^+_1}|$,
$\hat{T}^+_2\equiv \hat{T}_3^+\times
\vec{s}_\mu/|\hat{T}_3^+\times \vec{s}_\mu|$ and
$\hat{T}^+_1\equiv \hat{T}^+_2\times \hat{T}^+_3$. In the rest
frame  of the muon and in this coordinate system, \ba
P_{\mu^+}=(m_\mu,0,0,0)& \ \ \ & P_{e^-}=E_{e}(1,\sin\alpha \cos
\phi_-,\sin\alpha \sin \phi_-,\cos \alpha)\ \ \ \ \ \cr
P_{e^+_1}=(E_{e^+_1},0,0,E_{e^+_1})& \ \ \ &
P_{e_2^+}=(m_\mu-E_{e_1^+}-E_e,-E_{e}\sin\alpha \cos
\phi_-,-E_{e}\sin\alpha \sin \phi_-,-E_{e^+_1}-E_{e}\cos \alpha)
\ea where $\alpha$ is given by Eq.~(\ref{alpha}).

 Summing over the spins of $e^+_2$ and $e^-$, we
find   \ba \! &~& (2\pi)^3 (2E_e)(2E_{e^+_2}) \int
\sum_{\vec{s}_{e^-},\vec{s}_{e^+_2}}|M|^2 d\phi_-\cr \! &=&
(|B_1|^2 |c_{e^+_1}|^2+|B_2|^2|d_{e^+_1}|^2)E_e
[m_\mu-E_{e^+_1}(1-\cos
\alpha)]+(|B_1|^2|d_{e^+_1}|^2+|B_2|^2|c_{e^+_1}|^2)(1-\cos
\alpha)E_e(m_\mu-E_e-E_{e^+_1}) \cr \! &+&
 16(|G_2|^2|c_{e^+_1}|^2+|G_1|^2|d_{e^+_1}|^2)E_e(m_\mu-E_e(1-\cos
 \alpha))
 \cr
  \! &+& (|C_1|^2|c_{e^+_1}|^2+|C_2|^2|d_{e^+_1}|^2)E_e(m_\mu-E_e(1-\cos
 \alpha)) ~~~~~~~~
 \cr
 \!
&+&\mathbb{P}_\mu\left[(|B_1|^2|c_{e^+_1}|^2-|B_2|^2|d_{e^+_1}|^2)
[m_\mu-E_{e^+_1}(1-\cos \alpha)]  \cos \theta_+ \right. \cr \! &+&
\left. \mathbb{P}_\mu
(|B_2|^2|c_{e^+_1}|^2-|B_1|^2|d_{e^+_1}|^2)(1-\cos\alpha)(E_e
\cos\alpha+E_{e^+_1})\right] E_e \cos\theta_+ \cr\!
&+&\mathbb{P}_\mu(|B_2|^2|c_{e^+_1}|^2-|B_1|^2|d_{e^+_1}|^2)
\cos\phi_-E_e^2\sin\alpha(1-\cos\alpha) \sin\theta_+~~~
 \cr
 \! &+& \mathbb{P}_\mu(|C_1|^2|c_{e^+_1}|^2-|C_2|^2|d_{e^+_1}|^2)E_e\cos \alpha
 (m_\mu-E_e(1-\cos
 \alpha))\cos\theta_+
  ~~~~~~~~\cr
\! &+&
 16\mathbb{P}_\mu(|G_2|^2|c_{e^+_1}|^2-|G_1|^2|d_{e^+_1}|^2)E_e(m_\mu-E_e(1-\cos
 \alpha))\cos \alpha \cos \theta_+
 \cr
  \! &+&\mathbb{P}_\mu {
 \rm Re}[B_1C_2^*c_{e^+_1}d_{e^+_1}^*]m_\mu E_e
 \sin\theta_+(1+\cos\alpha)
 \cr
 \! &+& \mathbb{P}_\mu{
 \rm Re}[B_2C_1^*c_{e^+_1}^*d_{e^+_1}]m_\mu E_e
 \sin\theta_+(1+\cos\alpha)
\cr
 \! &+& 4\mathbb{P}_\mu\left[{
 \rm Re}[G_1B_1^*c_{e^+_1}^* d_{e^+_1}]-{
 \rm Re}[G_2B_2^*d_{e^+_1}^* c_{e^+_1}]\right] E_e
 (m_\mu-E_e(1-\cos \alpha))(1-\cos \alpha)\sin \theta_+
  \label{MM+}
 \ea
 where  $(c_{e^+_1} \ d_{e^+_1})$  determines the spin of the emitted positron
 [see
Eq.~(\ref{spin-positron})].
 The differential  rate of the decay into a positron in a
direction that makes an angle of $\theta_+$ with the spin of the
initial muon is $$\sum_{\vec{s}_{e^-},
\vec{s}_{e^+_2}}\frac{d\Gamma(\mu^+\to
e^+_1e^-e^+_2)}{d\Omega}=\int_0^{2\pi}
\int_0^{m_\mu/2}\int_{m_\mu/2-E_{e^+_1}}^{m_\mu/2}
\sum_{\vec{s}_{e^-},\vec{s}_{e^+_2}} |M|^2 E_eE_{e^+_1}(m_\mu -E_e
-E_{e^+_1}) dE_e~dE_{e^+_1} d\phi_- \  $$
 where $d\Omega$
is the differential solid angle determining the orientation of
$e^+_1$.
 Inserting $|M|^2$ from Eq.~(\ref{MM}), we obtain
\ba &~&\sum_{\vec{s}_{e^-}, \vec{s}_{e^+_2}}{d\Gamma(\mu^+\to
e^+_1e^-e^+_2)\over d\Omega}\cr &=&{m_\mu^5\over
8(2\pi)^4}\left[(|d_{e^+_1}|^2|B_1|^2+|c_{e^+_1}|^2|B_2|^2)
\int_0^{1/2}\int_{1/2-y}^{1/2} (1-x-y)(2x+2y-1)dx~dy\right. \cr
&+&(|c_{e^+_1}|^2|B_1|^2+|d_{e^+_1}|^2|B_2|^2)
\int_0^{1/2}\int_{1/2-y}^{1/2} y(1-2y)~dx~dy \cr
&+&(|C_1|^2|c_{e^+_1}|^2+|C_2|^2|d_{e^+_1}|^2)
\int_0^{1/2}\int_{1/2-y}^{1/2}x(1-2x) dx~ dy\cr &+&16
(|G_2|^2|c_{e^+_1}|^2+|G_1|^2|d_{e^+_1}|^2)\int_0^{1/2}\int_{1/2-y}^{1/2}
x(1-2x) dx~dy \cr &+&\mathbb{P}_\mu\cos \theta_+
(|B_1|^2|d_{e^+_1}|^2-|B_2|^2|c_{e^+_1}|^2)
\int_0^{1/2}\int_{1/2-y}^{1/2}(2x+2y-1)(-y
-\frac{1-2x-2y+2xy}{2y})] dx~ dy\cr &+&\mathbb{P}_\mu\cos
\theta_+(|C_1|^2|c_{e^+_1}|^2-|C_2|^2|d_{e^+_1}|^2)
\int_0^{1/2}\int_{1/2-y}^{1/2}(1-2x){1-2x-2y+2xy\over 2y} dx~
dy\cr &+&\mathbb{P}_\mu\cos \theta_+
(|B_1|^2|c_{e^+_1}|^2-|B_2|^2|d_{e^+_1}|^2)
\int_0^{1/2}\int_{1/2-y}^{1/2}y(1-2y) dx~ dy\cr  &+&
16\mathbb{P}_\mu\cos
\theta_+(|G_2|^2|c_{e^+_1}|^2-|G_1|^2|d_{e^+_1}|^2)\int_0^{1/2}\int_{1/2-y}^{1/2}
(1-2x){1-2x-2y+2xy \over 2y}dx~dy \cr
 &+&2\mathbb{P}_\mu({\rm
Re}[B_1C_2^*c_{e^+_1}d_{e^+_1}^*] +{\rm
Re}[B_2C_1^*c_{e^+_1}^*d_{e^+_1}]) \sin
\theta_+\int_0^{1/2}\int_{1/2-y}^{1/2}(2xy+0.5-x-y) dx~ dy \cr &+&
4 \mathbb{P}_\mu({\rm Re}[G_1B_1^* c_{e^+_1}^* d_{e^+_1}]-{\rm
Re}[G_2B_2^* c_{e^+_1} d_{e^+_1}^*]) \sin \theta_+
\int_0^{1/2}\int_{1/2-y}^{1/2} {(1-2x)(2x+2y-1)\over 4 x^2}dx~dy\
.\ea
\begin{acknowledgments}
I would like to thank M. Arjang for useful discussions. I am
especially grateful to M. M. Sheikh-Jabbari for his enriching
remarks. I would also like to thank the anonymous referee for
his/her useful remarks.
\end{acknowledgments}

\end{document}